\documentclass[sigconf,10pt]{acmart}
\renewcommand\footnotetextcopyrightpermission[1]{}

\usepackage{titlesec}
\usepackage{graphicx}
\usepackage{subcaption}

\titlespacing*{\section}{0pt}{6pt plus 4pt minus 2pt}{2pt plus 2pt minus 2pt}
\titlespacing*{\subsection}{0pt}{4pt plus 2pt minus 1pt}{2pt plus 1pt minus 1pt}
\titlespacing*{\subsubsection}{0pt}{4pt plus 2pt minus 1pt}{2pt plus 1pt minus 1pt}
\usepackage{tikz}
\usepackage{xcolor}
\usepackage{amsmath}

\usepackage{graphicx, epstopdf, stfloats, bbding, capt-of}

\usepackage{algorithm}
\usepackage{algorithmic}
\usepackage{enumitem }
\usepackage{multirow, multicol, booktabs, tabulary, tabu, longtable, array, varwidth}
\usepackage{placeins, lipsum}
\usepackage[flushleft]{threeparttable}
\usepackage{booktabs} %
\usepackage{makecell} %
\usepackage{listings}
\usepackage{fancyvrb}
\usepackage{ulem}
\usepackage{pifont}
\newcommand{\cmark}{\textcolor{green}{\ding{51}}} %
\newcommand{\xmark}{\textcolor{red}{\ding{55}}} %
\newcommand{\parab}[1]{\vspace{0.5\baselineskip}\noindent\textbf{#1}~}

\newcommand{\eg}{{\it e.g.,}\xspace}

\newcounter{BONumberOfComments}
\stepcounter{BONumberOfComments}

\newcounter{BeitongNumberOfComments}
\stepcounter{BeitongNumberOfComments}

\newcounter{LingzhiNumberOfComments}
\stepcounter{LingzhiNumberOfComments}

\newcounter{ShanboNumberOfComments}
\stepcounter{ShanboNumberOfComments}

\newcounter{HaozhenNumberOfComments}
\stepcounter{HaozhenNumberOfComments}

\newcounter{MingyuanNumberOfComments}
\stepcounter{MingyuanNumberOfComments}

\usepackage{cleveref}
\crefformat{section}{\S#2#1#3} %
\crefformat{subsection}{\S#2#1#3}
\crefformat{subsubsection}{\S#2#1#3}

\usepackage{xspace}
\newcommand{\sysname}{$\sf{AquaVLM}$\xspace}

\makeatletter
\renewcommand{\@titlefont}{\LARGE\sffamily\bfseries}
\makeatother

\title{\sysname: Improving Underwater Situation Awareness with Mobile Vision Language Models}

\newcommand{\colspace}{\hspace{20pt}}

\author{
  \Large
  \begin{tabular}{c @{\colspace} c @{\colspace} c @{\colspace} c}
    Beitong Tian$^*$ & Lingzhi Zhao$^*$ & Bo Chen &  Haozhen Zheng \\
    Jingcheng Yang & Mingyuan Wu & Deepak Vasisht & Klara Nahrstedt
  \end{tabular}
}
\renewcommand{\authors}{Beitong Tian, Lingzhi Zhao, Bo Chen, Haozhen Zheng,  Jingcheng Yang, Mingyuan Wu, Deepak Vasisht, Klara Nahrstedt}
\renewcommand{\shortauthors}{B. Tian, L. Zhao, B. Chen, H. Zheng, J. Yang, M. Wu, D. Vasisht, K. Nahrstedt}

\affiliation{
  \vspace{8pt}
  \institution{\Large University of Illinois Urbana-Champaign}
  \country{}
  \vspace{10pt}
}

\thanks{$^*$Both authors contributed equally to this work.}

\begin{document}

\begin{abstract}
Underwater activities like scuba diving enable millions annually to explore marine environments for recreation and scientific research. Maintaining situational awareness and effective communication are essential for diver safety.
Traditional underwater communication systems are often bulky and expensive, limiting their accessibility to divers of all levels. While recent systems leverage lightweight smartphones and support text messaging, the messages are predefined and thus restrict context-specific communication. 

In this paper, we present \sysname, a tap-and-send underwater communication system that automatically generates context-aware messages and transmits them using ubiquitous smartphones.
Our system features a mobile vision-language model (VLM) fine-tuned on an auto-generated underwater conversation dataset and employs a hierarchical message generation pipeline. We co-design the VLM and 
transmission, incorporating error-resilient fine-tuning to improve the system's robustness to transmission errors. 
We develop a VR simulator to enable users to experience \sysname in a realistic underwater environment and create a fully functional prototype on the iOS platform for real-world experiments. 
Both subjective and objective evaluations validate the effectiveness of \sysname and highlight its potential for personal underwater communication as well as broader mobile VLM applications.

\end{abstract}

\maketitle

\vspace{-10pt}
\section{Introduction}
\label{sec:introduction}

\begin{figure}[t]
\includegraphics[width=\linewidth]{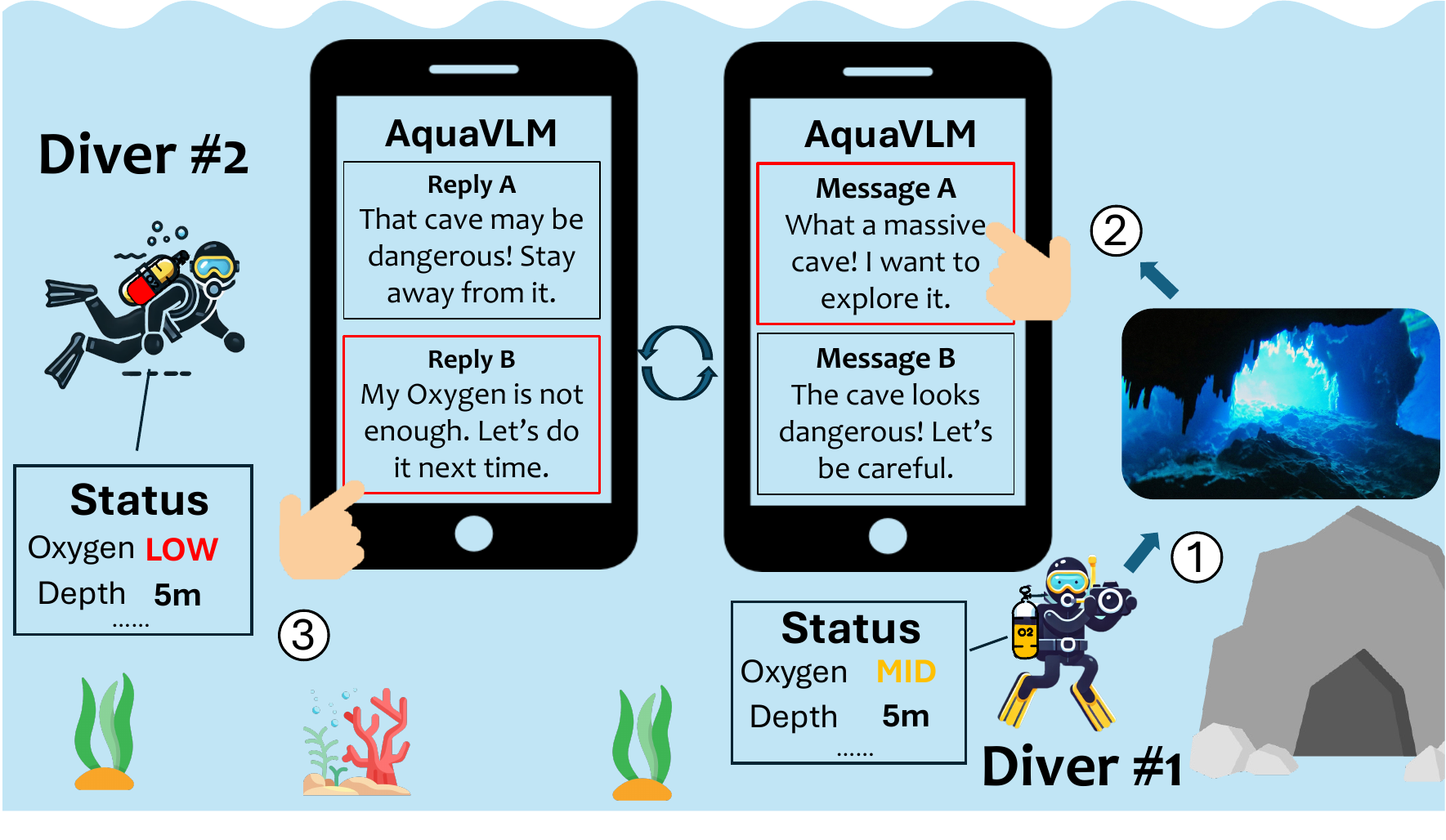}
\caption{Context-rich information generated and shared by \sysname.}
\label{fig:conceptual_graph}
\end{figure}

Underwater activities like scuba diving are the primary means for humans to explore the underwater world, with tens of millions of people engaging in these activities annually for various purposes such as recreation and scientific research~\cite{stat_zhao_plus,stat_zhao,stat_zhao_1,stat_zhao_2,stat_zhao_3}. Situation awareness is critical for enhancing diver safety and communication as it helps recognize potential danger and coordinate diving plans among divers~\cite{Sale2015,padi_instruction}.

Traditionally, scuba divers monitor critical information such as oxygen levels and dive duration through their diving computers~\cite{garmin_descent_t2_2023,GarminDiveComputers}. They communicate their diving statuses, such as "Out of air" or "Ascend", to dive buddies, via hand gestures~\cite{hand_signal,hand_signal_2}, light signals~\cite{lin2021shrimp} or acoustic signals~\cite{yang2023aquahelper,chen2022underwater,chen2023underwater,GarminUnderwaterCommunication, SuuntoWirelessTransmitter,garmin_descent_t2_2023}. However, these messages are generally predefined and do not provide detailed perceptions of the surrounding underwater environment. As a result, the limited informational content may negatively affect divers' judgment and decision-making during dives. While these limitations can be addressed by commercial underwater communication devices~\cite{DIVINTOK2024, OTS_BuddyPhone, OceanReefGSMGDivers} that enables direct talking like walkie-talkie, and integrating multimodal sensors~\cite{TKDE11,JOE22} for better perception of the environment, these technologies are typically \textit{bulky and expensive}
and therefore not widely accessible or ubiquitously adopted among recreational divers.

Considering the limitations of the above methods, we pose a question: \textit{Can divers share context-rich information underwater to improve situational awareness effortlessly with their mobile devices?} To achieve this, divers should consider and process various information such as the surrounding environment and diving status. Recent advancements in vision language models (VLMs) offer a promising solution~\cite{openai_gpt4o,Qwen2VL,chu2023mobilevlm}. VLMs have demonstrated remarkable capabilities in understanding multimodal data, including images and textual information, and generating coherent conversations. 
By deploying VLMs on mobile devices, we can accurately interpret real-world diving scenarios and assist divers by generating and sharing underwater-specific messages.

In this paper, we introduce {\sysname}, a novel underwater communication system powered by VLMs. {\sysname} is designed to facilitate informative messaging via mobile devices, as illustrated in Fig.~\ref{fig:conceptual_graph}. 
At a high level, the mobile VLM on the device analyzes \textit{multimodal data}, including images captured by divers and sensor readings from their mobile devices, to assess the current diving situation and generate suitable message options. Divers can select a message, which is then transmitted via acoustic signals to nearby diving partners.
On the receiving end, the VLM processes the incoming message along with contextual information to generate relevant candidate responses. Recipients can select and send a proper reply. 
Unlike traditional methods that rely on predefined messages or cumbersome techniques, {\sysname} allows divers to share detailed and context-rich underwater information to improve situation-awareness with less effort.

\vspace{-4pt}
\parab{Challenges.} Although the aforementioned design seems intuitive, its practical effectiveness faces two challenges.
First, \textit{adapting mobile VLM to underwater scenarios for generating informative messages is non-trivial}.
Mobile VLMs lack the generalization capabilities and domain-specific knowledge required to generate meaningful underwater conversations without fine-tuning. However, such fine-tuning is heavily dependent on large-scale, high-quality multimodal datasets, which are currently unavailable.
Additionally, constraints such as limited screen size and computational latency significantly restrict the quantity of generated messages. Consequently, ensuring the generation of satisfactory messages within these constraints becomes crucial.
Second, \textit{accurately sharing messages is equally crucial for effective situational awareness}. Incorrect or misunderstood messages may lead to inappropriate responses or actions, potentially placing both divers in dangerous situations. However, existing VLMs assume that all input information from the sender is transmitted accurately. In reality, underwater transmission is prone to errors which impacts message reliability.

\vspace{-4pt}
\parab{Context-aware instruction tuning (\cref{sec:instruction-tuning} \& \cref{subsec:hierarchical_message_generation}).} To generate effective underwater conversations, we instruct-tune an existing mobile VLM using a custom multimodal dataset tailored to diving scenarios.
Initially, we utilize a commercial VLM (e.g., ChatGPT-4o) to generate diverse underwater conversations, organizing these into three subtasks—message generation, message recovery, and reply generation—along with corresponding prompts. This dataset of instruction-answer pairs is used for fine-tuning the mobile VLM. 
Additionally, we incorporate different purposes—such as safety alerts and environmental descriptions—directly into the prompts. By specifying these purposes, the mobile VLM produces fewer, yet relevant, message options in a hierarchical manner, thereby reducing computational overhead.

\vspace{-4pt}
\parab{Error-resilient fine-tuning (\cref{subsubsec:fault_tolerant_vlm_finetuning}).}
To ensure reliable message exchange among divers, we further fine-tune the mobile VLM on underwater datasets with randomly disturbed messages to improve the system's robustness. 
Specifically, given the inherent language modeling capabilities of VLMs, we transmit the message character by character and directly apply the transmission errors aligned with the bit error rate (BER) collected from the real world to characters which facilitates VLM to recover messages with a certain degree of character corruption like humans.

To evaluate the effectiveness of \sysname, we design and build a virtual reality (VR)-based simulation platform that enables users to wear a headset to explore an underwater world, experience various events, and communicate with virtual divers using \sysname at any point during the simulation.
Our subjective evaluation demonstrates the effectiveness of the instruction-tuning pipeline with a hierarchical message generation process, achieving an 80\% purpose-align rate.
We also develop a fully functional prototype system on the iOS platform to evaluate the reliability of \sysname. 
Real-world experiments show that our system consistently maintains an average 90\% similarity between the received and original messages over distances of up to 20 meters.

Our key contributions are as follows:
\begin{itemize}[leftmargin=1em]
\item We design and build \sysname, the first underwater communication system leveraging mobile VLMs to enhance underwater situational awareness.
\item \sysname employs context-aware instruction tuning based on multimodal data, substantially improving its comprehension and adaptability within underwater environments.
\item \sysname implements error-resilient fine-tuning to improve the reliability of message transmission among divers.
\item We comprehensively evaluate \sysname through both subjective and objective metrics using an immersive VR evaluation platform and an iOS prototype, respectively, demonstrating its effectiveness and potential to advance underwater technologies.
\end{itemize}

\section{Background and Motivations}
\label{sec:background}
In this section, we first motivate the use of mobile devices for underwater communication and highlight the challenges of their adoption. We then introduce a new interaction paradigm that leverages mobile VLMs to enable more informative underwater messaging.

\begin{table*}[ht]
\centering
\caption{Comparisons of personal underwater communication methods. (Ubi. = Ubiquitous. ~ Info. = Informative)}
\begin{minipage}{0.5\textwidth} %
    \small
    \begin{tabular}{@{}p{4cm}p{3cm}p{1.2cm}p{1.2cm}c c@{}}
    \toprule
    \textbf{Device Type} &
    \textbf{Example} &
    \textbf{Price} &
    \textbf{Distance} &
    \textbf{Ubi.} &
    \textbf{Info.} \\ \midrule

    (1) Device-Free & 
    Hand signal~\cite{hand_signal_2,ParagonDiveGroup2024}& 
    \$0-\$50 & 
    <10m & 
    \cmark & 
    \xmark \\ \midrule

    (2) Underwater Talking Device & 
    Buddy Phone~\cite{OTS_BuddyPhone} & 
    \ > \$1000 & 
    \textgreater 50 m & 
    \xmark & 
    \cmark \\ \midrule

    (3) Messaging Diving Computer & 
Garmin~\cite{garmin_descent_t2_2023},Suunto~\cite{SuuntoD5BlackLime}& 
    \ > \$1500 & 
    <10 m & 
    \xmark & 
    \xmark \\ \midrule

    (4) Mobile Phone-based App & 
    AquaApp~\cite{chen2022underwater} &
    \$0-\$50 & 
    <30m & 
    \cmark & 
    \xmark \\ 
    \midrule

    Mobile Phone-based App (Ours) & 
    \sysname & 
    \$0-\$50 & 
    <20 m& 
    \cmark & 
    \cmark \\

    \bottomrule
    \end{tabular}
\end{minipage}%
\hfill
\begin{minipage}{0.25\textwidth} %
    \centering
    \begin{minipage}{0.48\textwidth}
        \centering
        \includegraphics[width=2.2cm]{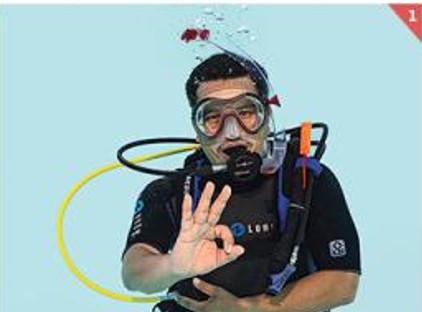}\\
        {\small (1)~\cite{scubadiving_hand_signals_image}}
    \end{minipage}%
    \hfill
    \begin{minipage}{0.48\textwidth}
        \centering
        \includegraphics[width=2.2cm]{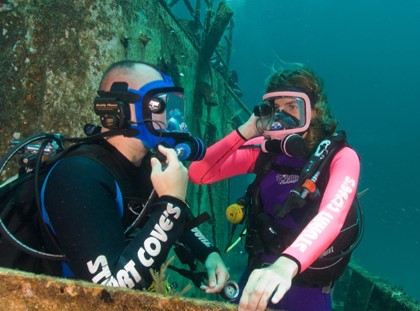}\\
        {\small (2)~\cite{buddy_phone_image}}
    \end{minipage}\\
    \begin{minipage}{0.48\textwidth}
        \centering
        \includegraphics[width=2.2cm]{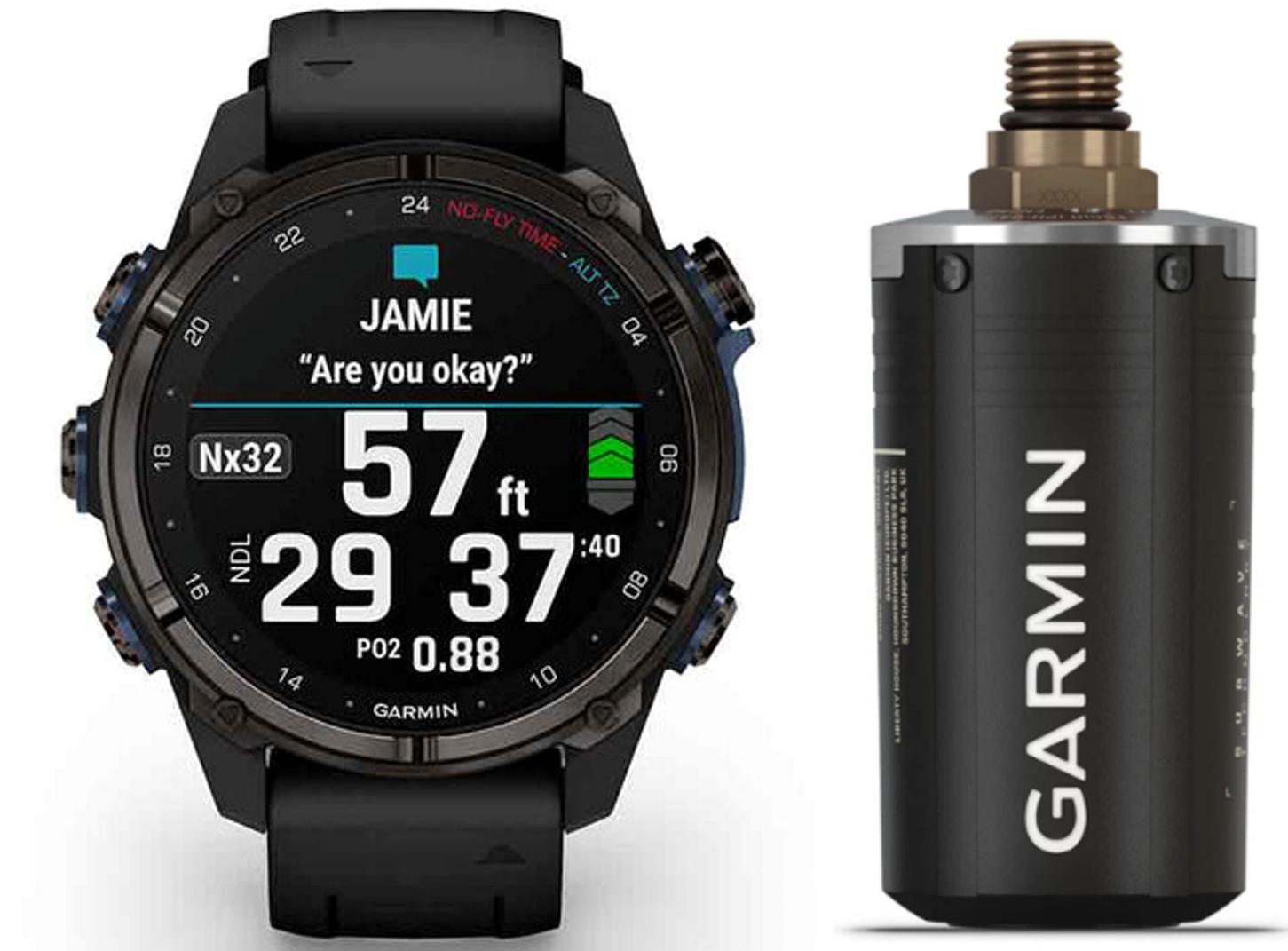}\\
        {\small (3)~\cite{garmin_descent_t2_2023}}
    \end{minipage}%
    \hfill
    \begin{minipage}{0.48\textwidth}
        \centering
        \includegraphics[width=2.2cm]{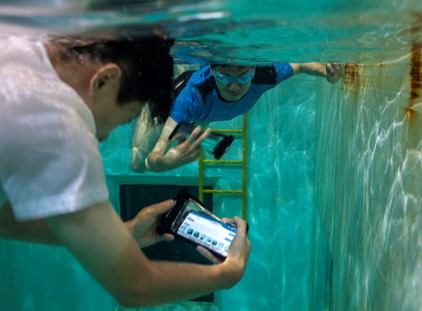}\\
        {\small (4)~\cite{uw_aquaapp_image}}
    \end{minipage}
\end{minipage}
\label{tab:personal_underwater_communication}
\end{table*}

\subsection{Personal Underwater Communication}
A personal underwater communication system enables multiple divers to exchange information underwater and adjust the diving plan.
Existing methods and products can be categorized into four groups as summarized in Table~\ref{tab:personal_underwater_communication}: 
\begin{enumerate}[leftmargin=*]
 \item Device-free. Traditional methods like hand and flash signals rely on predefined gestures (10-200) and clear visibility, rendering them ineffective in murky conditions or over longer distances. 
  \item Underwater talking device. Piezoelectric-based transceiver systems, such as Buddy Phone~\cite{OTS_BuddyPhone} and DIVINTOK~\cite{DIVINTOK2024}, facilitate direct verbal communication beyond 50m. However, these solutions are expensive (\$1000), bulky, and require intrusive equipment (e.g., full-face masks), restricting widespread usage.
 \item Diving computer with messaging features. Devices like Garmin~\cite{garmin_descent_t2_2023} and Suunto~\cite{SuuntoD5BlackLime} allow divers to exchange simple preset messages within 10 meters. Despite being compact, their high cost ($\sim$1500) and limited messaging capacity constrain effective communication.
 \item Mobile phone-based app. 
 Recent studies~\cite{chen2022underwater, chen2023underwater,yang2023aquahelper} leverage acoustic communication via smartphones to deliver predefined messages underwater. Given smartphones' widespread availability, computational capability, and integrated sensors, this method offers a versatile, accessible, and cost-effective solution. Consequently, \sysname is developed as a smartphone application.
 \end{enumerate}

\vspace{-4pt}
\parab{Challenges on mobile devices.} Operating a smartphone underwater—typically enclosed in a waterproof pouch differs fundamentally from usage in air, as conventional interaction methods such as typing or voice input become infeasible. Water disrupts the touchscreen’s sensitivity, reducing reliable interaction to single taps on large on-screen elements, while physical buttons are too few (two to four on most phones) to support efficient custom input. Consequently, divers are restricted to \textit{predefined messages}, which lack flexibility and cannot capture contextually meaningful communication shaped by their surroundings or sensor data. These challenges highlight the need for new underwater interaction paradigms. \sysname addresses this by leveraging VLMs on smartphones to interpret divers’ context and intent under such constrained conditions.

\subsection{Opportunity with VLM}

\vspace{-4pt}
\parab{Vision language models.} 
VLMs like ChatGPT4o~\cite{openai_gpt4o}, Qwen2-VL~\cite{Qwen2VL}, and MobileVLMV2~\linebreak[1]\cite{chu2024mobilevlm} are Transformer-based neural networks that contain billions of parameters and are pre-trained on massive amounts of multimodal data (\eg texts and images).
VLMs can respond to multimodal inputs with human-tone speeches, which achieve superior performance than prior neural network architectures on a wide spectrum of tasks (\eg image captioning, visual question answering, and visual grounding).
Although VLMs are known for their notoriously high computation and memory overheads, researchers have made tremendous efforts to run VLMs on mobile phones with model compression techniques such as model pruning, distillation, and quantization. For instance, a 4-bit quantized VLM with 3 billion parameters occupies approximately 4.5 GB of memory, making it feasible for mobile phones like the iPhone 14.
MobileVLMV2~\cite{chu2024mobilevlm} stands as the latest and open-source multimodal visual language model tailored for mobile devices.
The emergence of powerful multimodal understanding capabilities enabled by VLMs on mobile devices offers significant opportunities to enhance underwater situational awareness in two aspects.

\vspace{-4pt}
\parab{Contextual understanding and summarization.}  
First, VLMs enable mobile devices to understand and reason about the user's observations and status by integrating visual inputs with sensor data. This allows them to describe underwater contexts comprehensively by analyzing and summarizing multimodal signals.  

\vspace{-4pt}
\parab{Human-like language generation.} 
Second, VLMs can synthesize information from the user's environment and intent, generating human-like responses to provide more informative messages. When prompted appropriately, mobile VLMs produce messages tailored to diverse user intents, reflecting the varied corpus on which they are trained.

\vspace{-4pt}
\section{System Overview}
\label{sec:systemdesign_message_generator}

\begin{figure*}[t!]
    \centering
\includegraphics[width=\linewidth]{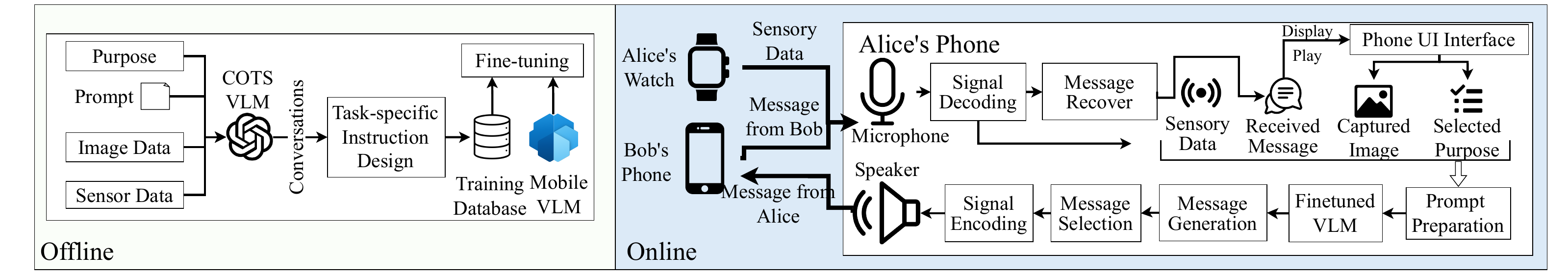}
    \caption{System overview and workflow of \sysname. In the offline stage (left), a customized VLM model is fine-tuned. In the online stage (right), the message is generated and transmitted between two mobile devices. }
    \label{fig:system_overview}
\end{figure*}

\begin{figure}[t]
    \centering
    \includegraphics[width=\linewidth]{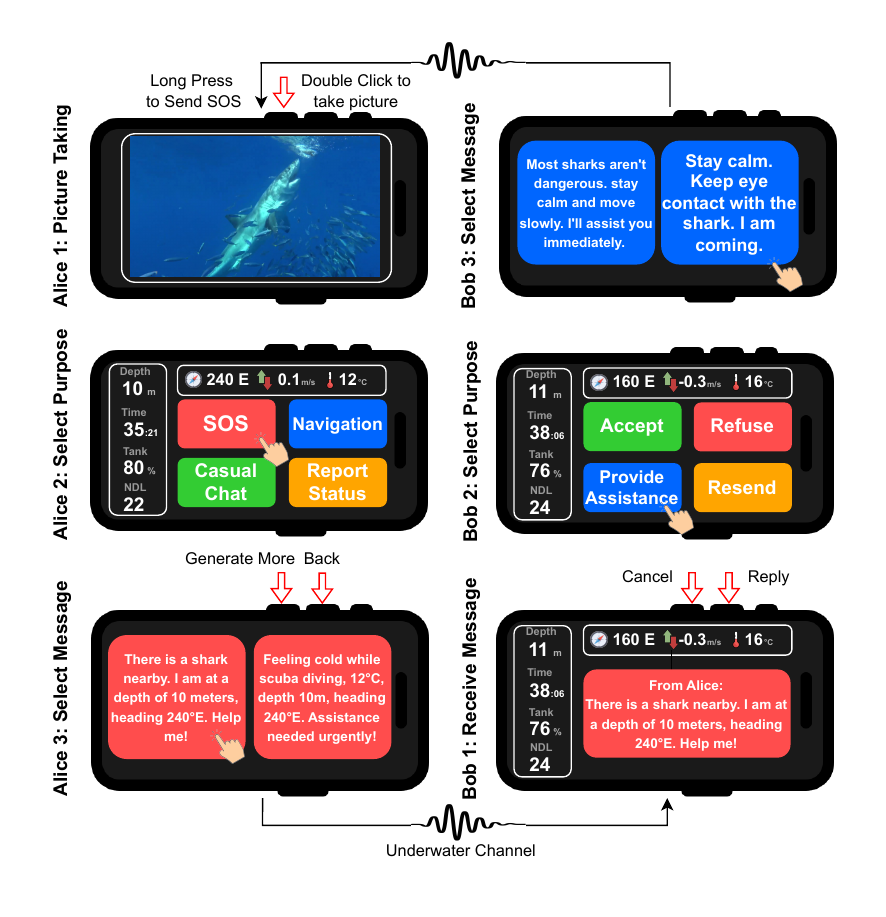}
    \vspace{-25pt}
    \caption{\sysname user interface with an example where divers encounter a shark.}
    \label{fig:example_UI}
\end{figure}

Fig.~\ref{fig:system_overview} provides an overview of \sysname. \sysname operates in two stages: offline and online. In the offline stage, the primary focus is on instruct-tuning the mobile VLM using automatically generated datasets which enhances its abilities in context understanding, conversation generation, and corrupted message recovery in underwater scenarios. 
During operation, the mobile VLM runs on each user's mobile phone to generate messages and replies. These are based on the user's purpose and available multimodal context, including images captured in real-time by the phone's camera, sensory data from the phone, and other wearable devices like diving watches. 
Once generated, the diver selects one of the messages, which is converted into a bitstream, encoded using channel coding, and subsequently modulated into acoustic signals for underwater transmission.
At the receiver side, the received signals are demodulated and decoded to reconstruct the message.
This reconstructed message is then passed to the mobile VLM for error recovery.
Finally, recovered messages are either displayed visually on the user's phone screen or converted into audio.
In Fig.~\ref{fig:example_UI}, we show the user interface with an example between Alice and Bob.

\vspace{-4pt}

\section{\sysname Design}
In this section, we introduce key designs of \sysname: context-aware instruction tuning and error-resilient fine-tuning.
\subsection{Context-aware Instruction Tuning }\label{sec:instruction-tuning}

State-of-the-art mobile VLMs~\cite{chu2023mobilevlm,chu2024mobilevlm} cannot directly generate underwater conversations due to limited generalizability, discrepancies between training and underwater datasets, and task-specific differences. To address these challenges, we adapt the mobile VLM to the underwater domain via a carefully designed instruction tuning. 

\vspace{-4pt}
\parab{Multimodal data preparation.} 
\label{subsubsec:underwater_conversation_preparation}
We extracted 1,368 keyframes from five scuba diving videos captured across diverse diving locations and scenarios~\cite{dataset_video_1,dataset_video_2,dataset_video_3,dataset_video_4,dataset_video_5}.
For sensor data, we selected nine critical parameters typically available on mobile devices or diving watches. Due to the absence of publicly available diving sensor data, we simulated typical recreational diving scenarios, including descent, exploration, ascent, safety stops, and dive completion, to synthesize representative sensor datasets.

\begin{figure*}[t!]
    \centering
    \includegraphics[width=\linewidth]{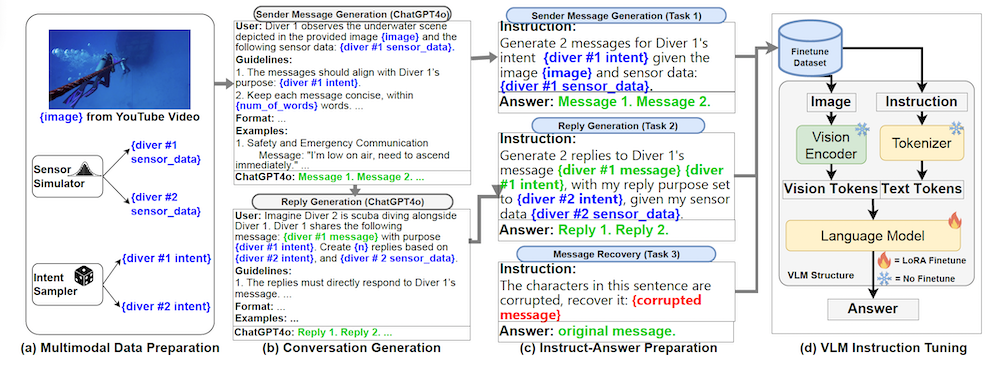}
    \vspace{-10pt}
    \caption{Model instruction tuning pipeline.}
    \label{fig:finetune_pipeline}
\end{figure*}

\vspace{-4pt}
\parab{Conversation generation.}
We generate underwater conversation using commercial VLMs, i.e., ChatGPT-4o, leveraging self-collected multimodal data and carefully designed prompts, as illustrated in Fig.~\ref{fig:finetune_pipeline} (a) and (b). Our prompts clearly described diving scenarios, provided concise guidelines (e.g., simple language, brief responses), and specified desired output formats. We also included illustrative few-shot examples within prompts to guide the VLM in generating precise and relevant outputs.

\vspace{-4pt}
\parab{Task definition.} We identify three tasks for instruction tuning, as illustrated in Fig.~\ref{fig:finetune_pipeline} (c): The first two tasks are \textit{sender message generation} and \textit{reply generation} which align with the two steps in the conversation generation step in Fig.~\ref{fig:finetune_pipeline} (b). For these tasks, we simplify the instruction prompt to retain only essential details, such as task descriptions, and multimodal data (e.g., images, sensor data, and user intent). The answers are the messages and replies extracted from the generated underwater conversation corpus by ChatGPT-4o. The number of messages and replies can vary from 1 to 4, depending on design requirements and system preferences. In our implementation, we use two messages to balance diversity and generation speed. We also define a \textit{message recovery} task to recover corrupted received messages during transmission which will be detailed further in \cref{subsubsec:fault_tolerant_vlm_finetuning}. As shown in Fig.~\ref{fig:finetune_pipeline} (c), instead of generating entire conversations simultaneously, we employ a two-step modular approach: first, generating initial messages based on the sender's purpose; second, crafting appropriate replies. This strategy offers several advantages:
first, by dividing the task into focused subtasks, our method aligns effectively with chain-of-thought prompting~\cite{wei2022chain}, enhancing reasoning and promoting diversity in multi-turn conversations;
second, modular generation reduces common formatting errors, simplifying data processing and improving dataset quality.

\vspace{-4pt}
\parab{Instruction tuning.}
After processing the underwater conversation corpus for each task and filtering out duplicated and less useful samples, we obtain an instruction tuning dataset comprising 16,000 samples for \textit{sender message generation}, 21,000 samples for \textit{reply generation}, and 48,000 samples for \textit{message recovery}. These samples are merged into a unified instruction-tuning dataset.
The instruction tuning process uses Low-Rank Adaptation (LoRA)~\linebreak[1]\cite{hu2021lora}
for efficiency and focuses exclusively on the language component of the mobile VLM, as shown in Fig.~\ref{fig:finetune_pipeline} (d). 
LoRA freezes pre-trained weights and introduces trainable, low-rank matrices in specific layers, reducing trainable parameters while enabling efficient task-specific fine-tuning.
During inference, the vision encoder converts images (if available) into vision tokens containing visual information. Simultaneously, user instructions and sensor data are tokenized into text tokens representing semantic meaning. The vision tokens and text tokens are concatenated and processed by the language model to generate accurate answers.

\vspace{-4pt}
\subsection{Hierarchical Message Generation}
\label{subsec:hierarchical_message_generation}
Mobile VLM, even after instruction tuning, still struggles to generate message candidates that precisely align with the user's intent using only image and sensor data. This limitation stems from its constrained model size. A straightforward solution to this issue is to generate a larger number of message candidates (\eg eight messages) to increase the likelihood of producing a user-satisfying option. While simple, this approach incurs a huge computational burden.

We address the above challenges via hierarchical message generation based on users' intents. Diver first selects a purpose, which then serves as part of a prompt to guide the mobile VLM in generating messages aligned with that specific intent. By narrowing the focus, this method reduces the number of generated messages from 8 to 2 while maintaining relevance to the user's needs. 
During instruction tuning (\cref{sec:instruction-tuning}), when generating the underwater conversation, we add randomly chosen intent in the prompt to guide ChatGPT-4o to generate messages aligned with the chosen intent, which allows the finetuned model to generalize to diverse intents. Specifically, we define four purposes for message generation, i.e., safety, navigation,  environment, and equipment. For reply generation, the purposes are: acknowledge, refuse, assist, and resend. These communication purpose categories can be adapted to suit the needs of specific user groups. 

During the inference, we find not all sensory data are necessary, and an excess of normal sensory data in the prompt can sometimes confuse the mobile VLM. This may hinder its ability to extract critical information from images and essential sensory inputs, as well as generate meaningful messages within two message candidates. To address this, we employ a threshold-based approach to retain only abnormal or potentially hazardous sensor readings in the prompt (e.g., water temperature below $15 ^\circ\mathrm{C}$ or tank pressure below 700 psi). This approach implicitly integrates general scuba diving knowledge into the system, enhancing the mobile VLM's ability to process and respond to relevant inputs.

\begin{figure}[t!]
    \centering
    \includegraphics[width=0.9\linewidth]{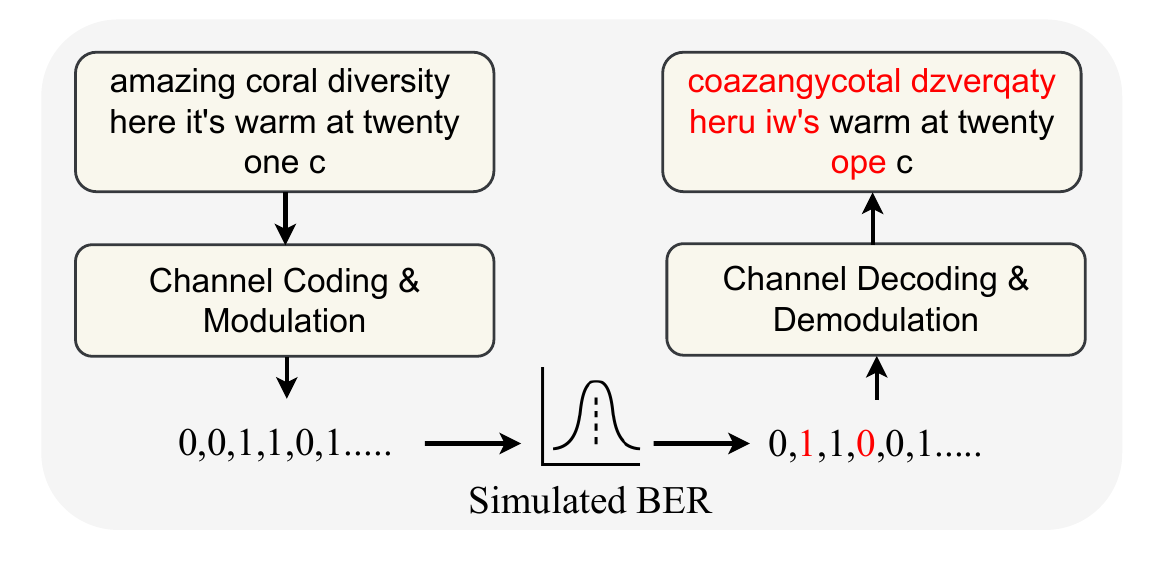}
    \caption{Error-resilient fine-tuning: Each bit of a character is randomly flipped given a BER.}
    \label{fig:fault_tolerant_vlm_fine_tuning}
\end{figure}

\subsection{Error-resilient Fine-tuning}
\label{subsubsec:fault_tolerant_vlm_finetuning}

The natural language modeling capabilities of VLMs offer a unique opportunity to reduce errors in received messages. 
For example, even when a prompt contains numerous typos, the VLM can often interpret the intended meaning and generate an appropriate response. This robust error-handling ability can be leveraged to recover corrupted messages in underwater communication, a potential unexplored in previous methods~\cite{chen2022underwater}.
To this end, \sysname\ transmits messages character by character. 
However, underwater transmission is highly susceptible to transmission errors caused by channel variations at the PHY layer. Even though channel coding~\cite{hamming1986coding} at PHY layer can handle transmission errors to some extent, its effectiveness is limited especially when the BER is large. To further improve the robustness of the system, we fine-tune the mobile VLMs using pairs of corrupted and original messages from the underwater dataset.

To simulate errors, we construct a transmission pipeline at PHY layer which includes encoding, modulation, decoding, etc. When converting the message to bitstreams, we optimize the efficiency and robustness by using a smaller number (i.e., 5) of bits and converting the numerical values into text (e.g., "12.1" to "twelve point one"), respectively.
At the PHY layer, we adopt the CSS (chirp spread spectrum) techniques~\cite{steinmetz2022taking,TSN23} to convert the messages into chirp signals since it provides a good balance between the robustness and data rate~\cite{steinmetz2022taking,xie2024icc,TSN23}. Based on our real-world measurement, the BER is selected from 0\% to 20\% in 1\% increments. Each selected BER will be applied to a randomly selected subset of sender and reply messages generated in~\cref{sec:instruction-tuning} (as shown in Fig.~\ref{fig:fault_tolerant_vlm_fine_tuning}), with randomly selected purposes prepended to each message. In total, we generate 48,000 simulated corrupted original message pairs, which are then combined with data from other tasks to fine-tune the model. While there are several loss-resilient transmission frameworks~\cite{nsdi24_grace,li2024reparolossresilientgenerativecodec,aquascope}, they mainly focus on fine-tuning the neural codecs for video or image delivery. Instead, \sysname is the first work to improve the error-resilient of VLMs for text transmissions.

\vspace{-4pt}
\section{Implementation}
\label{sec:implementation}

\begin{figure*}[t]
    \centering
    \begin{subfigure}[b]{0.33\textwidth}
    \includegraphics[width=\linewidth]{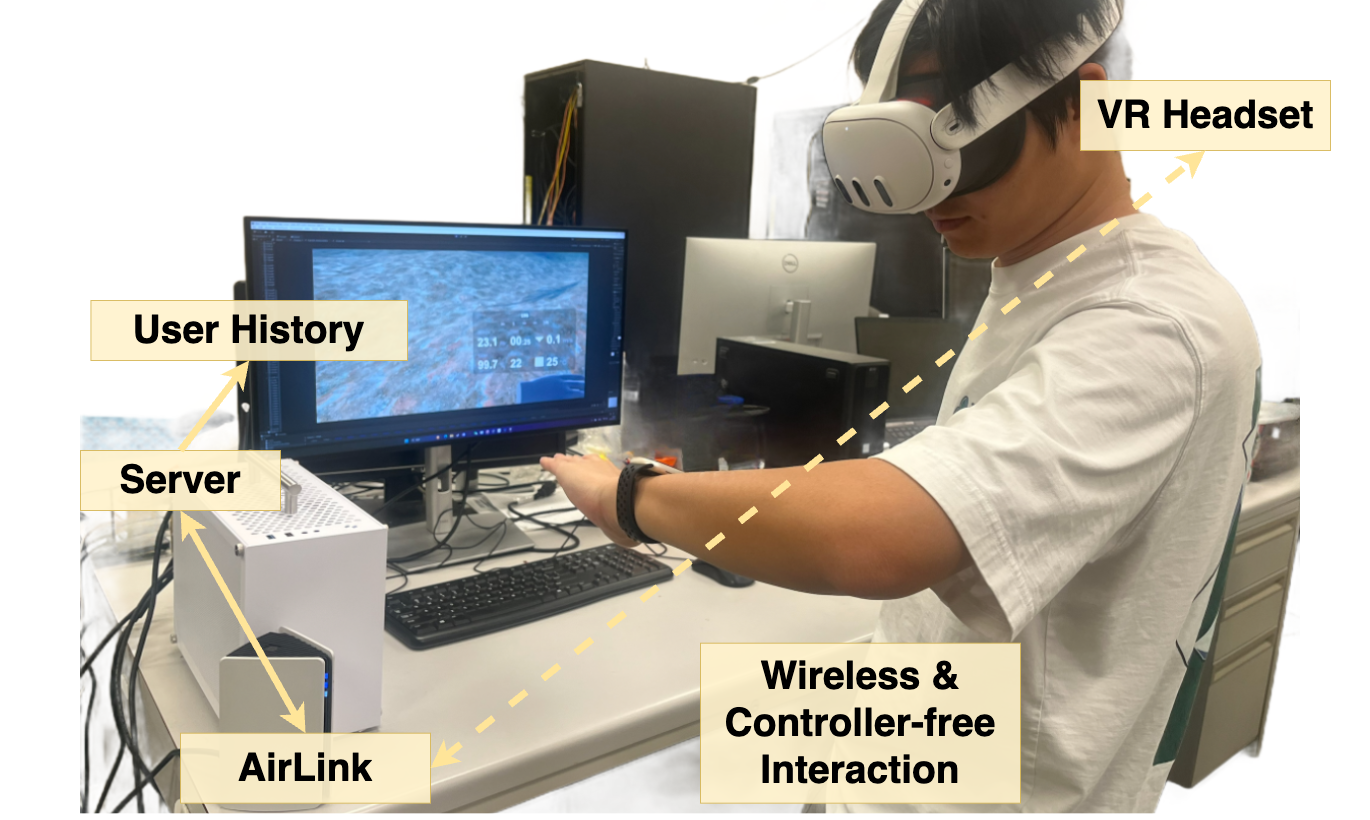}
      \caption{User study testbed.}
    \label{fig:testbed}
    \end{subfigure}
        \begin{subfigure}[b]{0.33\textwidth}
    \includegraphics[width=\linewidth]{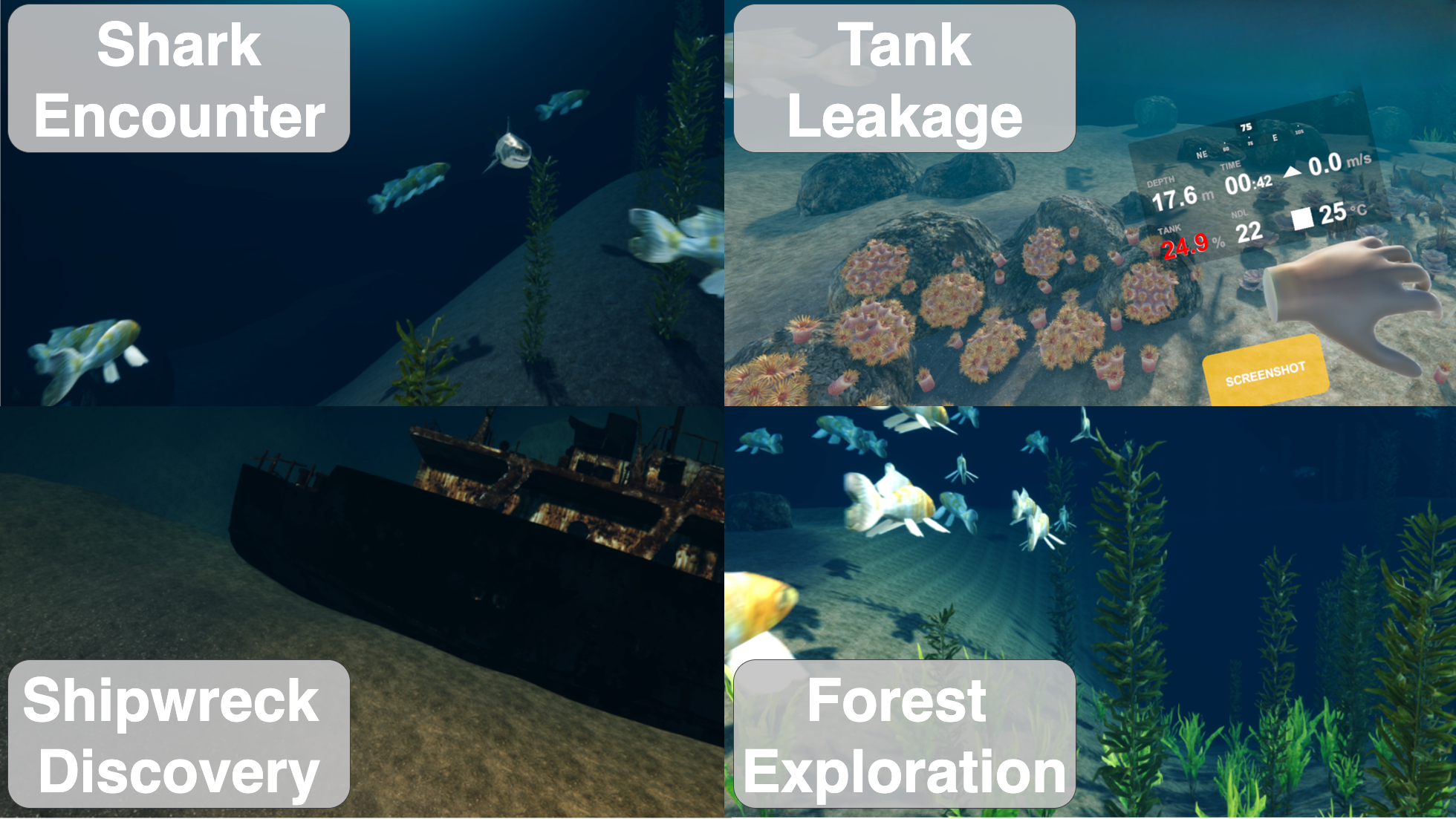}
      \caption{VR underwater events.}
\label{fig:snapshot}
    \end{subfigure}
    \begin{subfigure}[b]{0.33\textwidth}
    \includegraphics[width=\linewidth]{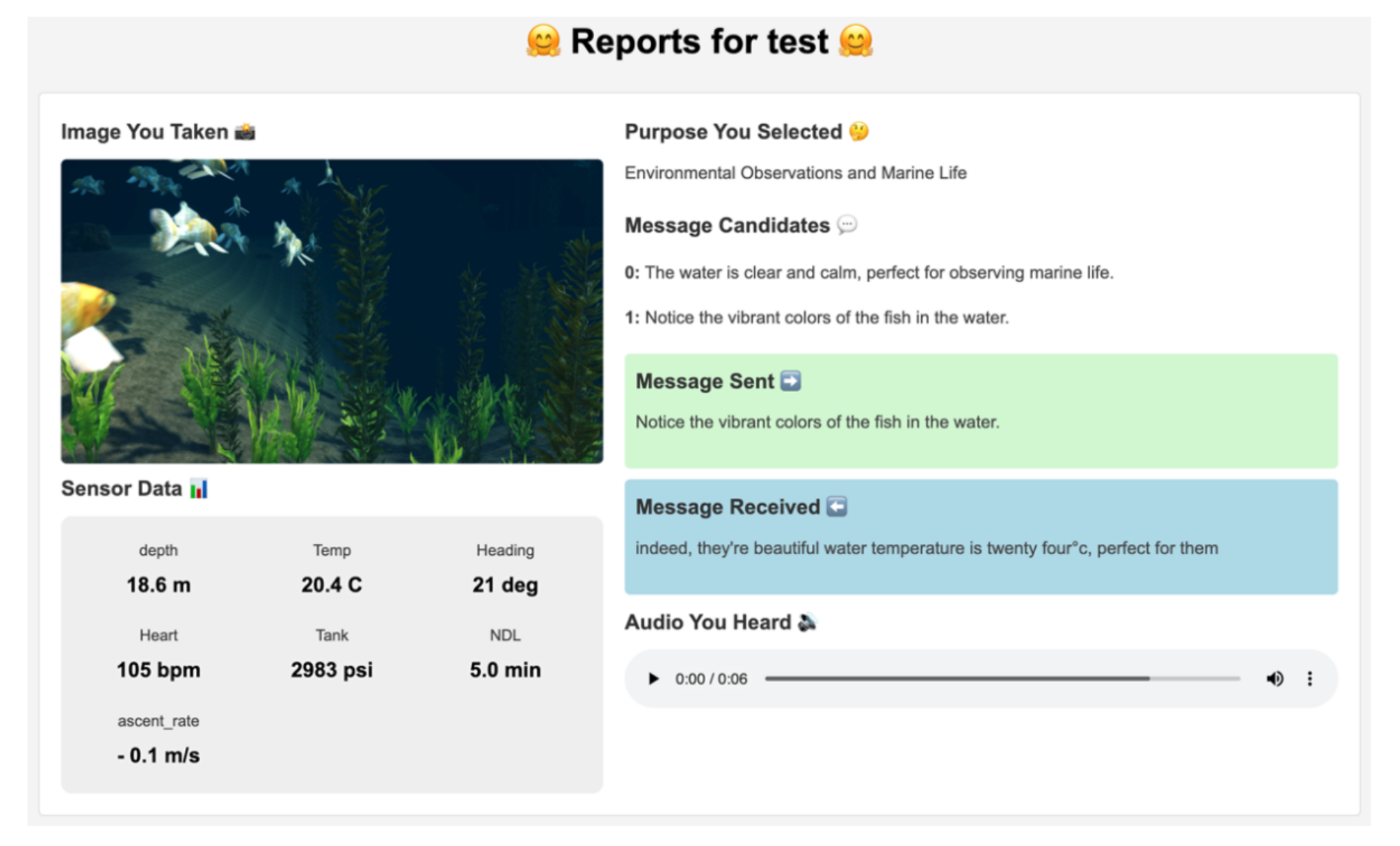}
      \caption{Website interface.}
\label{fig:website}
    \end{subfigure}
    \caption{User study setup.}
    \label{fig:user_study_setup}
\end{figure*}

\subsection{Model Training and Deployment}
We deploy MobileVLMV2~\cite{chu2024mobilevlm}, a state-of-the-art efficient VLM with 3B parameters. Fine-tuning uses LoRA with a rank of 64, a cosine learning rate schedule (warm-up ratio: 0.03, starting at $4 \times 10^{-5}$), and the AdamW optimizer~\cite{adamw} to minimize overfitting and forgetting.
The model is fine-tuned on an 85k-entry dataset with a batch size of 16 per GPU, running for one epoch on two NVIDIA RTX 4090 GPUs in 4 hours.
For the VR platform, the fine-tuned model is quantized to 4 bits for efficient mobile inference using llama.cpp~\cite{gerganov_llama_cpp}, a C/C++ implementation optimized for diverse hardware. The model is converted to GGUF format and deployed via the LLMFarm interface~\cite{guinmoon_llmfarm} to iOS.

\subsection{VR Simulation Platform}\label{sec:simulated_VR_platform}
We developed a Unity-based underwater conversation simulator for the Meta Quest 3, allowing users to experience \sysname in a realistic virtual environment. 
Fig.~\ref{fig:testbed} shows the testbed, with user view snapshots in Fig.~\ref{fig:snapshot}.

\vspace{-4pt}
\parab{Diving events.} 
We simulate a realistic VR environment using photorealistic 3D assets and real underwater soundtracks to provide a fully immersive visual and auditory underwater experience.
To enhance engagement and realism, we have designed four dynamic scenarios: (1) \textit{Shark Encounter}, where a shark slowly swims toward the diver; (2) \textit{Tank Leakage}, which simulates a critical equipment malfunction; (3) \textit{Shipwreck Discovery}, an exciting find of a sunken vessel; and (4) \textit{Kelp Forest Exploration}, a peaceful experience through a dense kelp forest designed for casual chat and marine life discovery. To add variety and unpredictability, these events are triggered randomly during user testing, ensuring a diverse and immersive simulated experience.
To tackle the low frame rates commonly associated with headset-based simulations, which can cause dizziness and poor user experiences, we render the frames on the backend server and stream them to the headset via AirLink. 
The Airlink ensures a cable-free, high-FPS experience that enhances immersion.

\vspace{-4pt}
\parab{User interaction.} 
We realize the interaction with the user interface of \sysname by creating a virtual phone screen in the user's viewport. 
The image captured by the user is sent to the backend server that runs the same model deployed on the mobile phone to derive the answers. 
Then, the answers are sent back to the headset.
To ensure a realistic experience with \sysname, we emulate delays with real-world captured latency traces with mobile phones.
Furthermore, we develop a hand gesture recognition algorithm that allows users to move by performing "swimming" motions, eliminating the need for traditional controllers. 
Besides, we log the interaction data and store them on the backend server which could facilitate future customization of \sysname.

\subsection{iOS Prototype}

We developed a prototype using an iPhone 12 Pro and Apple Watch Ultra (Fig.~\ref{fig:ios_prototype}). The model runs on the phone's GPU via Metal API~\cite{apple_metal}, with the app built using Swift and SwiftUI, and Accelerate lib for signal processing. The Apple Watch transmits sensor data (e.g., temperature, heart rate, depth~\cite{apple_submersion_data}) to the phone using acoustic signals. Images and no-decompression limits (NDL) are processed on the phone, while the Watch's proximity (within 1m to the iPhone) ensures reliable transmission despite its weaker speaker.
The phone is placed in a touch-sensitive waterproof pouch~\cite{TORRAS} rated for 100 feet, while Apple Watch Ultra is water-resistant to 130+ feet. The system can also be adapted for Android.

\vspace{-4pt}
\parab{Transmission pipeline.} The data bits of the message are encoded via Hamming encoding~\cite{hamming1986coding} at a $4/7$ code rate. We adopt the CSS modulation technique where each data symbol is converted to a chirp signal that spans the entire bandwidth~\cite{TSN23,jung2021exploiting}. The chirp generation process is governed by two key parameters: the spreading factor ($SF$) and the bandwidth ($BW$)~\cite{steinmetz2022taking}.
To balance robustness and data rate, we set $SF$ $=$ 5, achieving a data rate of 312 bps with $BW$ $=$ 2 kHz in underwater environments. 
We insert training symbols every 3 data symbols for time-equlization. We add a preamble at the head of the entire data packet to facilitate packet detection~\cite{chen2022underwater}. At the receiver side, its microphone continuously listens, recording sound for packet detection~\cite{chen2022underwater}. Detected packets are processed through synchronization, equalization, demodulation, and decoding. Received bits are converted to messages, with errors corrected by a VLM-based recovery module as detailed in \cref{subsubsec:fault_tolerant_vlm_finetuning}.

\vspace{-4pt}
\section{Evaluation}
\label{sec:evaluation}

\subsection{User Study Setup}
We conducted a user study with 20 participants (13 male, 7 female), aged 18–35. All participants provided informed consent, adhering to institutional ethics policies. %
First, participants completed a 5-minute tutorial to familiarize themselves with the VR platform and {\sysname}. They were briefed on key questionnaire topics for later evaluation. Then, using a VR headset, participants explored an underwater environment where 2–3 events were randomly triggered (\cref{sec:simulated_VR_platform}). Participants used \sysname\ to communicate with a simulated diver (played by us) to resolve events. Next, participants reviewed their session data, including images, sensor data, messages, and replies, via a webpage (see Fig.~\ref{fig:website}). They assessed whether messages aligned with intended purposes. Finally, participants completed a questionnaire with nine usability and experience questions rated on a 1–5 MOS (mean score opinion) scale, with 1 being the worst and 5 the best. Open-ended feedback was also collected. The total session lasts around 30 minutes.

\begin{figure}[t]
    \centering
    \begin{subfigure}[b]{0.23\textwidth}
    \includegraphics[width=\linewidth]{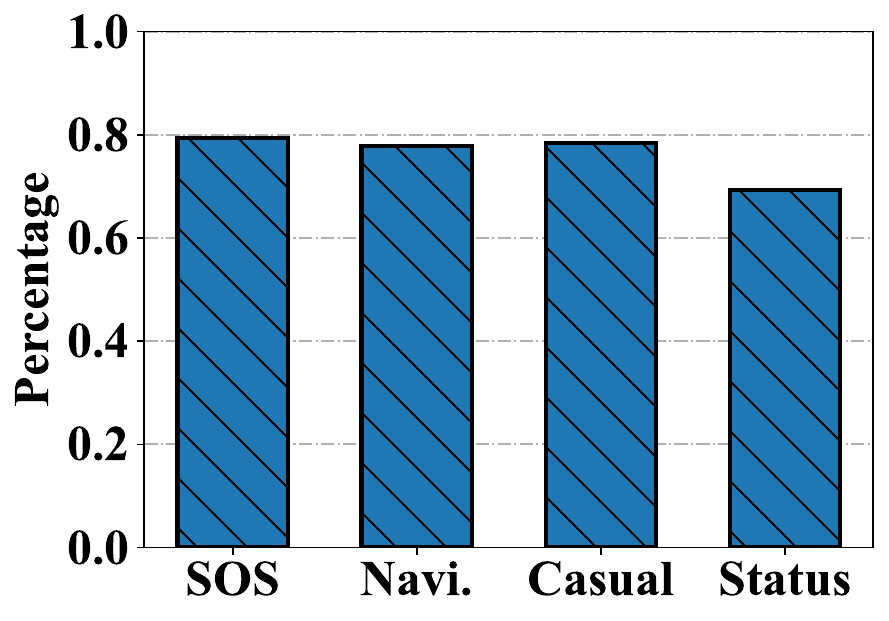}
      \caption{Purpose-align rate.}
    \label{fig:subjective_rate}
    \end{subfigure}
        \begin{subfigure}[b]{0.23\textwidth}
    \includegraphics[width=\linewidth]{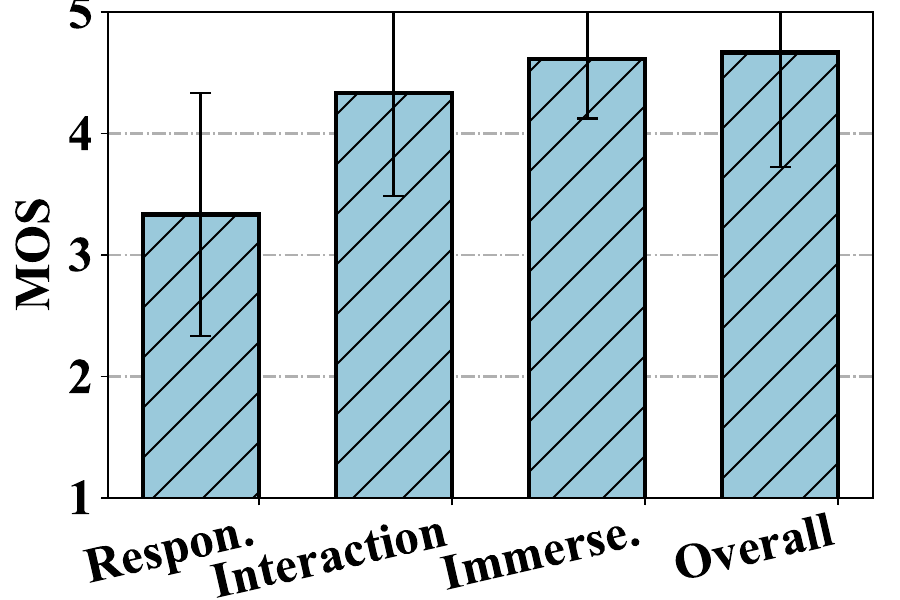}
      \caption{VR experiences.}
    \label{fig:subjective_results}
    \end{subfigure}
    \caption{User study results. (a) The generated messages by \sysname align well with users' purposes. (b) Users have a satisfactory overall VR experience.}
    \label{fig:user_study_results}
\end{figure}

\vspace{-4pt}
\subsection{Subjective Results}
\vspace{-4pt}
\parab{Metrics.} We consider five subjective metrics:
\begin{itemize}[leftmargin=*]
    \item \textit{Purpose-align Rate} is calculated as the ratio of purpose-aligned messages recognized by participants to the total number of generated messages in the user study. 
    \item \textit{Responsiveness} measures the impact of communication latency on participants' experience.
    \item \textit{Interaction} measures the intuitiveness of the interaction with the environments and generating the messages.
    \item \textit{Immersiveness} measures how realistic the simulated VR environments are. 
    \item \textit{Overall } measures the overall experience of using the VR platform and {\sysname}.
\end{itemize}

\vspace{-4pt}
\parab{Purpose-align rate.} The purpose-align rate ranges from 70\% to 80\% as shown in Fig.~\ref{fig:subjective_rate}, demonstrating that the VLM is capable of effectively generating purpose-aligned messages based on the provided intent and multimodal data in an underwater scenario. 
The casual purpose is the most frequently selected and achieves the high purpose-alignment rate because the mobile VLM consistently recognizes objects in images and initiates conversations about them. In contrast, the navigation purpose is rarely used and status purpose has the lowest alignment rate. Discussions with participants reveal that {\sysname} effectively handles scenarios like navigating to objects shown in pictures or using compass directions. However, it struggles when the destination is unclear, such as when participants are lost and want to return. In such cases, the mobile VLM fails to provide satisfactory answers. This limitation could be addressed by adding a "backtrace" purpose.
We emphasize that this valuable feedback is due to the realistic environment provided by the VR evaluation platform, which helps identify corner cases that might otherwise be difficult to discover. This allows such issues to be addressed early in the development process.
Some failures occur when users select the SOS purpose in non-emergency situations. Consequently, \sysname cannot derive useful information and generates two random SOS messages, misaligned with user intent. Another issue arises when objects in images are too small to be recognized, such as a distant shark. This limitation stems from the vision encoder's capabilities.
In summary, {\sysname} effectively handles cases where user intent aligns with the provided multimodal data and intent candidates but struggles when intent-related information is missing. In the future, {\sysname} can use observations from user study to iterate to get more intelligent.

\vspace{-4pt}
\parab{Participant experience.} As shown in Fig.~\ref{fig:subjective_results}, participants rate the interaction, immersiveness, and overall experience as satisfactory. However, the MOS for responsiveness is relatively low. This diminished responsiveness is attributed to the nature of the messaging system, where the receiver must interpret, compose, and send a response, leading to noticeable latency. Many participants initially expected the system to function like a telephone, blurring the distinction between a messaging system and a real-time voice communication system. After clarifying this concept, most participants found the latency acceptable.

\begin{figure}[t]
    \centering
    \begin{subfigure}[b]{0.23\textwidth}
    \includegraphics[width=\linewidth]{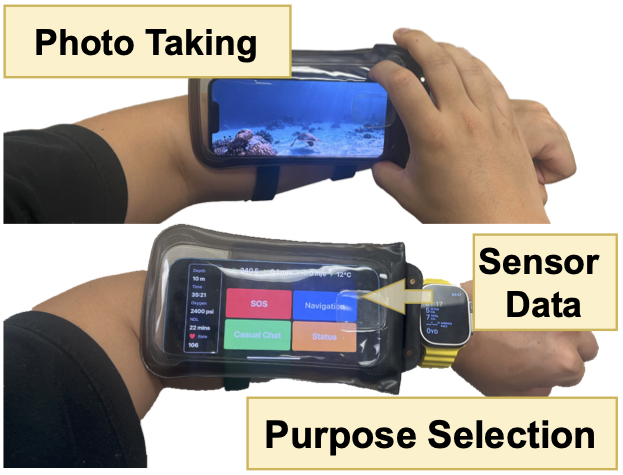}
      \caption{iOS prototype.}
    \label{fig:ios_prototype}
    \end{subfigure}
        \begin{subfigure}[b]{0.23\textwidth}
    \includegraphics[width=\linewidth]{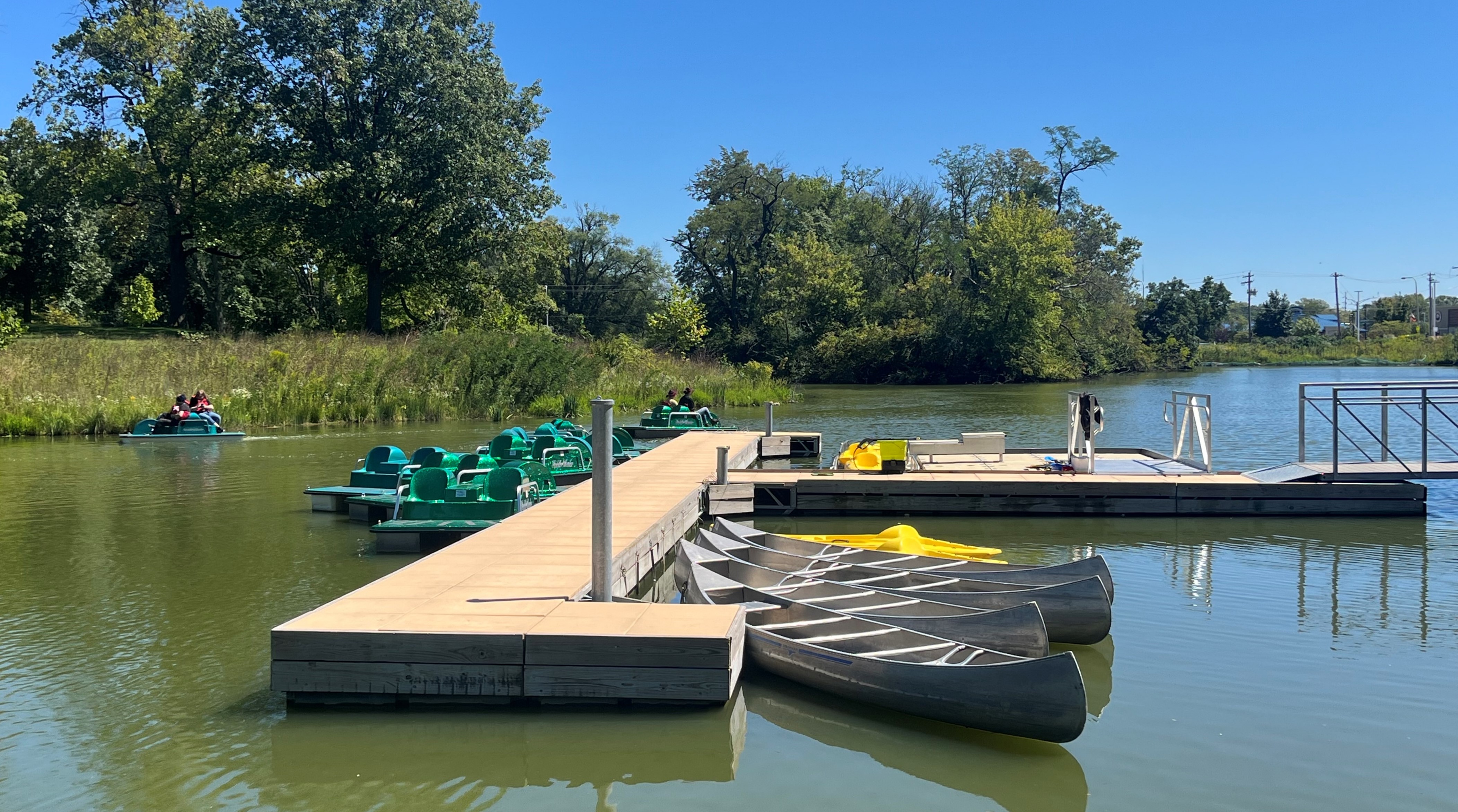}
      \caption{Test environment.}
\label{fig:environment}
    \end{subfigure}
    \caption{Real-world experiment setup.}
    \label{fig:real_world_setup}
\end{figure}

\begin{figure*}[t]
    \centering
    \begin{subfigure}[b]{0.24\textwidth}
    \includegraphics[width=\linewidth]{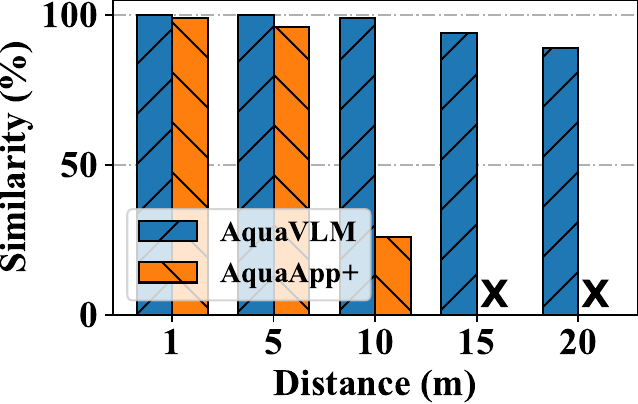}
      \caption{Distance \textit{vs.} similarity.}
    \label{fig:eval_distance_similarity}
    \end{subfigure}
        \begin{subfigure}[b]{0.24\textwidth}
    \includegraphics[width=\linewidth]{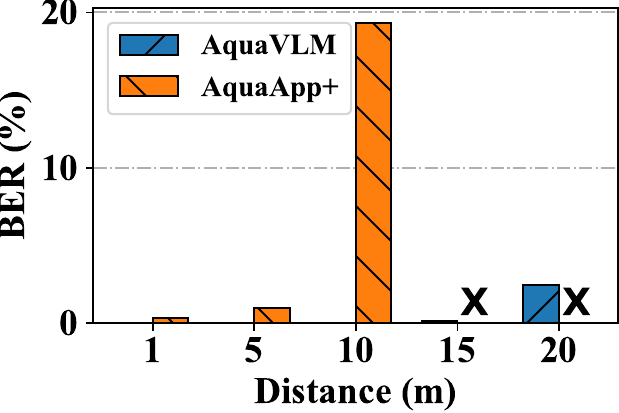}
      \caption{Distance \textit{vs.} BER.}
    \label{fig:eval_distance_ber}
    \end{subfigure}
        \begin{subfigure}[b]{0.24\textwidth}
    \includegraphics[width=\linewidth]{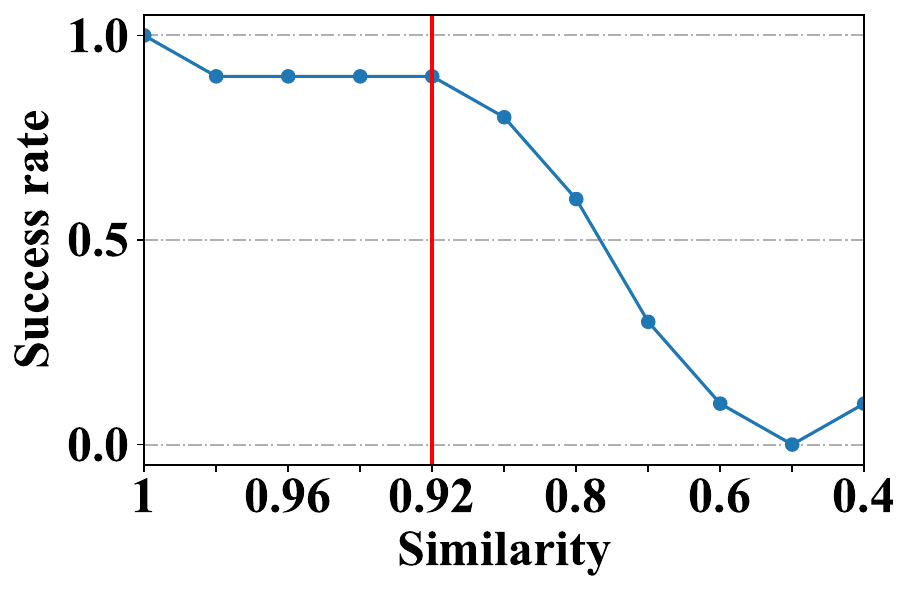}
      \caption{Similarity \textit{vs.} success rate.}
    \label{fig:success_rate_vs_similarity}
    \end{subfigure}
    \begin{subfigure}[b]{0.24\textwidth}
    \includegraphics[width=\linewidth]{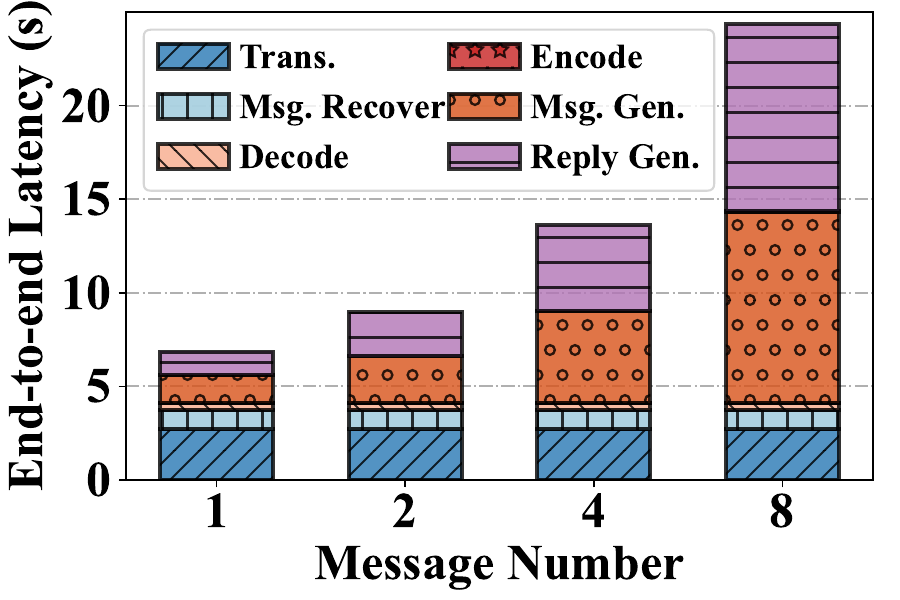}
      \caption{End-to-end Latency.}
    \label{fig:latency_breakdown}
    \end{subfigure}
    \caption{Real-world performance. (a) \sysname achieves consistently high semantic similarity while AquaApp+ fails to deliver meaningful messages when the transmission distance is over 10m. In (a) and (b), an ``X'' mark indicates that the receiver does not receive the message packets. (c) Similarity over 0.92 is recognized as semantically "same" by humans. (d) Transmission and inference times are the primary contributors to the end-to-end latency.}
    \label{fig:real_wolrd_performance}
\end{figure*}

\vspace{-4pt}
\subsection{Results in the Wild}

\vspace{-4pt}
\parab{Real world test setup.}
As shown in Fig.~\ref{fig:environment}, we conduct tests in a lake with a maximum range of 20m and an average depth of 3m. The sender's speaker and receiver's microphone are aligned at a depth of 2m. We transmit 100 conversational rounds at varying distances. Note that there is no existing mobile systems that support underwater communication with customized messages. We construct a baseline scheme, \textit{AquaApp+~\cite{chen2022underwater}}. AquaApp is originally designed for text transmission using a three-step OFDM protocol. Specifically, the sender transmits a preamble, the receiver responds with high-SNR subcarrier information, and the sender then encodes and sends the payload using those subcarriers. We mainly compare our system with the transmission protocol part of AquaApp. We integrate AquaApp systems with our VLM model to make it comparable to \sysname.

\vspace{-4pt}
\parab{BER and semantic similarity.}
We evaluate \sysname's ability to transmit long messages under various distances. We consider two metrics, BER and semantic similarity between the transmitted and recovered messages. The similarity is calculated using the cosine similarity of sentence embeddings from all-MiniLM-L6-v2~\cite{wang2020minilm}. This provides a more intuitive and comprehensible metric for determining whether the recovered message successfully conveys the meaning of the original message.

Fig.~\ref{fig:eval_distance_similarity} and Fig.~\ref{fig:eval_distance_ber} illustrate the transmission results where \sysname achieves low BER (<3\%) within 20m and the semantic similarity is over 90\% within 15m. Note that \sysname is fully implemented on a mobile device, which operates at much lower transmission power than specialized hardware. Achieving such a low BER under these constraints already represents state-of-the-art performance. However, BER of AquaApp+ surges to 20\% when the transmission distance is 10m, leading to a rapid drop in similarity. AquaApp+ suffers from transmission errors due to its suboptimal carrier selection for long packets and thus has a larger BER.

To investigate how the similarity reflects human perception of the received message, we conduct another user study. Specifically, we build a dataset of 200 recovered-original message pairs with semantic similarity values falling within a specific range (\eg 0.75--0.85). Participants are asked to judge whether the recovered messages are semantically "same" or "different" from the original messages in the context of underwater communication. We define \textit{success rate} as the fraction of semantically "same" messages. To improve accuracy, we segment similarity into multiple ranges from 0.4 to 1.0, with finer granularity between 0.9 and 1.0. As shown in Fig.~\ref{fig:success_rate_vs_similarity}, the success rate begins to drop significantly when similarity falls below 92\%. Based on this observation, we set the semantic similarity threshold at 92\%. Therefore, we can see that the received message by \sysname is semantically meaningful to the human when the transmission range is between 0-15m.
Furthermore, we evaluate its performance under varying depths, orientations, and motion patterns. The results show that changes in depth and orientation have minimal impact, and \sysname maintains a low BER (<2\%) even when subjected to accelerations of 2–2.5 $m/s^{2}$.

\vspace{-4pt}
\parab{One-round latency.}
Fig.~\ref{fig:latency_breakdown} summarizes the latency for a single communication round and its breakdown across message generation, recovery, reply generation, and transmission. The primary contributors to latency are VLM inference and transmission time, while physical layer computations like encoding and modulation are negligible. Latency increases significantly with more generated messages, so we default to generating two sender messages and replies, balancing efficiency and practicality. Since users send only one message per round, message recovery time remains constant.

\subsection{Ablation Study}
\begin{figure*}[t]
    \centering
    \begin{subfigure}[b]{0.33\textwidth}
    \includegraphics[width=\linewidth]{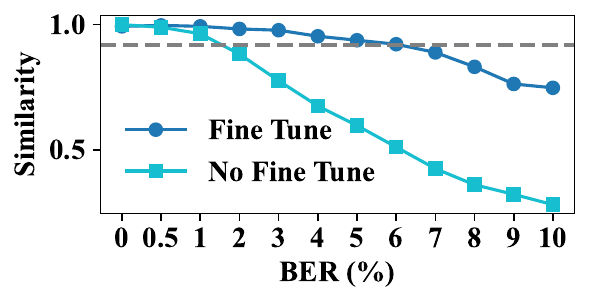}
      \vspace{-18pt}
      \caption{Impact of fine-tuning.}
    \label{fig:eval_fine_tune}
    \end{subfigure}
        \begin{subfigure}[b]{0.33\textwidth}
    \includegraphics[width=\linewidth]{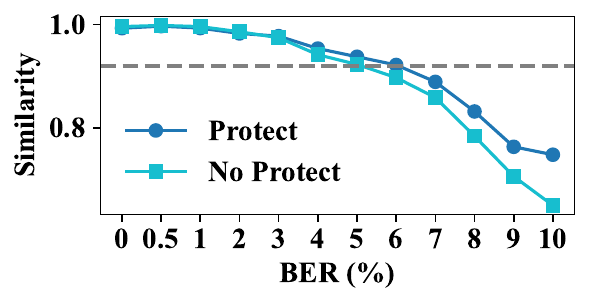}
      \vspace{-18pt}
      \caption{Impact of separator.}
    \label{fig:eval_separator}
    \end{subfigure}
    \begin{subfigure}[b]{0.33\textwidth}
    \includegraphics[width=\linewidth]{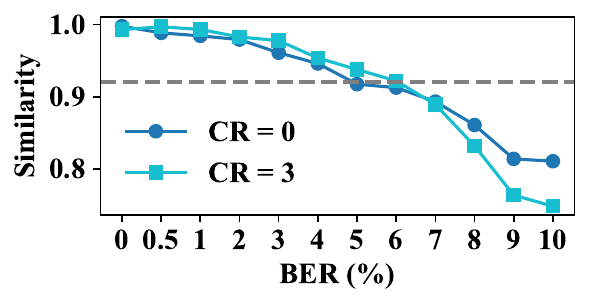}
       \vspace{-18pt}
      \caption{Impact of channel coding.}
    \label{fig:eval_channel_coding}
    \end{subfigure}
    \caption{Ablation Results. (a) \sysname maintains semantic meanings even when BER is at 6\%. (b) Spaces within the message have less impact on recovery capability. (c) Hamming coding's effectiveness diminishes with BER.}
    \label{fig:eval_ablation}
\end{figure*}

In this part, we investigate the impacts of fine-tuning, separator protection, and hamming coding on message recovery performance, as shown in Fig.~\ref{fig:eval_ablation}.
The gray dashed line shows the divider of semantically meaningful and meaningless.

\vspace{-4pt}
\parab{Impact of fine-tuning.}
Fig.~\ref{fig:eval_fine_tune} compares the performances of VLM with and without fine-tuning. 
As shown in Fig.~\ref{fig:eval_fine_tune}, our fine-tuned VLM that takes the transmission errors into account shows great power in recovering the corrupted message even when the BER is $6\%$ while the vanilla version starts to fail when BER is $1.5\%$. In addition, the similarity gain of fine-tuning can reach up to $114\%$.

\vspace{-4pt}
\parab{Impact of separator protection.}
We investigate whether the separator, specifically spaces within the message, can greatly contribute to the VLM's recovery capability. To evaluate this, we recover corrupted separators before inputting the message into the VLM.
Fig.~\ref{fig:eval_separator} presents the results, which differ from our initial expectations. Messages with separator protection demonstrate slightly higher similarity than those without, primarily due to a reduced error rate in the message. This counterintuitive result can be attributed to the auto-regressive generation process of the VLM, which operates differently from how humans process text.

\vspace{-4pt}
\parab{Impact of channel coding.}
Fig.~\ref{fig:eval_channel_coding} compares \sysname's performance with (CR = 3) and without (CR = 0) Hamming coding. For CR = 0, no physical-layer encoding is applied. For CR = 3, chirp-based encoding includes Gray coding, interleaving, and (7,4) Hamming coding. The VLM is fine-tuned with separators protected.
Even at a 6\% BER without Hamming coding, the VLM can recover meaningful messages, showing that an error-free channel isn't always required for text or image transmission. Instead, VLM recoverability can balance transmission reliability and efficiency.
Surprisingly, CR = 0 performs comparably to CR = 3 when BER < 7\% and even surpasses it when BER > 7\%. This is due to (7,4) Hamming coding's limitation to correct only one bit per four. As BER increases, Hamming coding's effectiveness diminishes, while the fine-tuned VLM reliably recovers messages at error rates below 5\%.

\vspace{-4pt}
\subsection{Rescource Usage}
We also evaluate the resource usage of mobile devices when running the mobile VLM. As shown in Table~\ref{tab:resource}, running the mobile VLM on an iPhone 12 Pro results in approximately 570\% CPU usage on its 6-core CPU, with GPU memory usage reaching 3.48 GB—an amount that modern smartphones, starting from the iPhone 14, can easily support. Battery consumption is moderate at 27\% per hour, which is significantly lower than the typical duration of a recreational dive (approximately one hour), ensuring sufficient operation time.

\begin{table}[t]
    \centering
    \small
    \caption{Resouce Usage}
    \begin{tabular}{ccccc}
        \toprule
           \textbf{Model Size}   & \textbf{CPU} & \textbf{GPU Memory} & \textbf{Battery Drop}  \\
        \midrule
        \textbf{MobileVLM-1.7B}  &  577\%   & 1.81 GB & 21\%/h  \\
        \textbf{MobileVLM-3B}  & 570\%   & 3.48 GB   & 27\%/h \\

        \bottomrule
    \end{tabular}%
    \label{tab:resource}
\end{table}

\vspace{-4pt}
\section{Discussion} \label{sec:discussion}
In this section, we discuss the future directions and potential
improvements of \sysname.

\vspace{-4pt}
\parab{Generalization and extension.} Although we focus specifically on underwater applications, we believe that several concepts introduced in this work are generalizable to other extreme environments where user input is constrained and enhanced situational awareness is required, such as firefighting. Our technique thus paves the way for similar advancements in other domains.

\vspace{-4pt}
\parab{Transmission distance.} \sysname's effective transmission distance is around 20m. To achieve longer distances, an attachable transceiver can be developed using cost-effective underwater modem designs~\cite{benson2010design, cario2015seamodem}. This is feasible on mobile devices, which allow upgrades, unlike fixed-interface commercial alternatives. 

\vspace{-4pt}
\parab{Latency improvement.} The large end-to-end latency mainly stems from model inference on mobile devices and the transmission delay. Fine-tuning smaller models (e.g., 1.7B parameters) could be a promising solution where it halves inference latency compared to 3B models. However, the current performance of the smaller model remains unsatisfactory. Future advancements in sub-3B models may address this limitation. To reduce the transmission delay, we can also leverage a lightweight underwater modem that provides wider bandwidth and thus higher data rate.

\vspace{-4pt}
\parab{Personalization.} 
VLM personalization improves user experience by tailoring outputs to preferences, similar to ChatGPT's "Memory"~\cite{openai_memory_controls_2024}. Data from the VR platform could fine-tune models for personalized diving assistance.

\vspace{-4pt}
\parab{User study feedback.} User feedback led to features like "Go back" and "Generate more candidates," improving the interaction loop. Participants praised the VR evaluation as "realistic," with immersive experiences contributing to valuable feedback for underwater communication and similar systems. To further enhance situational awareness, additional capabilities can be incorporated—such as detecting potentially dangerous, previously unseen creatures—by fine-tuning the VLMs on a diverse dataset of ocean life.

\vspace{-4pt}
\section{Related Work}
\label{sec:related_work}

\vspace{-4pt}
\parab{Ubiquitous underwater systems.} AquaApp~\cite{chen2022underwater} pioneers underwater text messaging between mobile phones via OFDM protocol, later adding 3D localization functionality~\cite{chen2023underwater}. AquaHelp~\cite{yang2023aquahelper,yang80neural} and ChirpCom~\cite{chirpcom} leverage chirp signal for underwater messaging. In particular, AquaHelp enhances SOS signal transmission with robust detection algorithms. ChirpCom~\cite{chirpcom} proposes a feedback mechanism to adaptively adjust SF based on the measured SNR from the receiver.
Similarly, QR-based methods allow divers to communicate by scanning the QR code on the smartphone~\cite{liu2021uqcom,liu2023uqrcom}. This approach relies on underwater visibility and distance. 

\sysname fundamentally differs from the existing work in two aspects: First, AquaVLM primarily aims to enhance divers’ situational awareness by generating and sharing context-specific informative messages, whereas previous work facilitates underwater texting by exchanging only pre-defined messages between divers. Second, AquaVLM focuses on employment and optimization of a VLM tailored for underwater-specific conversation. It is important to note that most of the optimizations occur at the APP layer. In contrast, most of the previous work focuses on robust transmission protocol designs at the PHY layer. Therefore, \sysname is orthogonal to previous underwater communication systems.

\vspace{-4pt}
\parab{Large model-based intelligence on mobile devices.}
Advancements in on-device intelligence by major tech companies emphasize integrating large models into mobile devices for automation. Apple offers features like rewriting and emoji generation \cite{apple_intelligence}, while systems like Autodroid~\cite{autodrio_mobicom24} and AutoGLM~\cite{zhipuai_official_website} enable voice-controlled operations. Smart glasses like Meta RayBan~\cite{meta_rayban_ai_features_2024} connect to servers for image-based queries. Unlike these, \sysname processes multimodal data entirely on-device, specializing in underwater communication, showcasing new possibilities for on-device intelligence in unique and challenging scenarios.

\vspace{-4pt}
\parab{Underwater situational awareness.}
Researchers~\cite{JOE22,TKDE11} have made significant strides in enhancing situational awareness for underwater vehicles (e.g., unmanned robots) to support tasks such as safe navigation and mission planning. To achieve this, they have developed specialized hardware equipped with multiple sensors—acoustic, vision, and depth sensors—to more effectively probe and understand the underwater environment. \sysname is the first work that improves situational awareness and enriches underwater communication for humans.

\section{Conclusion}
\label{sec:conclusion}

We introduce \sysname, the first underwater communication system for mobile phones, enabling divers to share detailed, contextual information effortlessly. \sysname achieved this by instruct-tuning a mobile VLM using a self-collected underwater dataset. Accessible to divers of all experience levels at almost no extra cost, \sysname only required a mobile app download. Both simulation and real-world evaluation results showed the effectiveness of \sysname and the potential of deploying large models on mobile devices to improve situational awareness.



\end{document}